\font\tx=cmr10 at 11pt \font\ma=cmmi10 at 11pt \font\sy=cmsy10 at 11pt
\textfont0=\tx         \textfont1=\ma          \textfont2=\sy
\font\sub=cmr8         \font\masub=cmmi8       \font\sysub=cmsy8
\scriptfont0=\sub      \scriptfont1=\masub     \scriptfont2=\sysub
\font\title=cmbx12     \font\author=cmr12      
\baselineskip=13pt     \parindent=18pt         \raggedbottom
\lefthyphenmin=3       \righthyphenmin=4       \hyphenpenalty=200
\def\space{\vskip 13pt}
\def\section#1{\space\space\goodbreak\centerline{#1}\nobreak\space}
\vglue 5mm
\tx
\hyphenation{recollapsing}
\centerline{\title  THE FORMATION OF THE FIRST STARS
         \footnote{$^1$}{\tx Presented at the 33rd ESLAB Symposium
        on ``Star Formation from the Small to the Large Scale'' held
         in Noordwijk, The Netherlands, November 2--5, 1999; to be
         published in the ESA Special Publications Series (SP-445),
              edited by F. Favata, A. A. Kaas, and A. Wilson}}
\space\space

\centerline{\author         Richard B. Larson}
\space
\centerline{            Yale Astronomy Department}
\centerline{          New Haven, CT 06520-8101, USA}
\centerline{               larson@astro.yale.edu}
\space\space\space

\centerline{                    ABSTRACT}
\space
{\narrower
   The first bound star-forming systems in the universe are predicted
to form at redshifts of about 30 and to have masses of the order of
$10^6\,$M$_\odot$.  Although their sizes and masses are similar to
those of present-day star-forming regions, their temperatures are
expected to be much higher because cooling is provided only by trace
amounts of molecular hydrogen.  Several recent simulations of the
collapse and fragmentation of primordial clouds have converged on a
thermal regime where the density is about $10^3$--$10^4\,$cm$^{-3}$
and the temperature is about 300$\,$K; under these conditions the
Jeans mass is of the order of $10^3\,$M$_\odot$, and all of the
simulations show the formation of clumps with masses of this order.
The temperatures in these clumps subsequently rise slowly as they
collapse, so little if any further fragmentation is expected.  As a
result, the formation of predominantly massive or very massive stars is
expected, and star formation with a normal present-day IMF seems very
unlikely.  The most massive early stars are expected to collapse to
black holes, and these in turn are predicted to end up concentrated near
the centers of present-day large galaxies.  Such black holes may play a
role in the origin of AGNs, and the heavy elements produced by somewhat
less massive stars also formed at early times may play an important role
in chemically enriching the inner parts of large galaxies and quasars.
\space}

\section{1.~~INTRODUCTION}

   How did star formation begin in the universe?  And can we make any
credible predictions about the properties of the first stars?  Recent
years have seen great advances in our theoretical understanding of the
origin of structure in the universe and the formation of galaxies, and
on the observational front we now have observations extending out to
redshifts greater than 5 and looking back to the first billion years of
the history of the universe.  It is clear that much had already happened
during that period: galaxies, or at least parts of galaxies, had already
appeared on the scene; the first quasars had already formed; and the
intergalactic medium had become ionized.  Furthermore, the densest parts
of the universe, including quasars, had already become significantly
enriched in heavy elements.  Most of the heavy elements in large
galaxies like our own appear in fact to have been produced at relatively
early times, and there has been only modest subsequent enrichment,
leaving a general paucity of metal-poor stars compared with the
predictions of simple models; this is the long-standing and apparently
ubiquitous `G-dwarf problem'.  Finally, a far-infrared background
radiation has been observed which contains half of the present radiative
energy density of the universe, and which is believed to have been
produced mainly by dust-obscured star formation at high redshifts.  All
of these observations reflect in various ways the effects of early star
formation, and some of them, particularly the rapid enrichment in heavy
elements, would be easiest to understand if early star formation had
produced preferentially massive stars. Therefore there has been great
interest in understanding the earliest stages of star formation in the
universe, and especially in understanding what the typical masses or
mass spectrum of the first stars might have been.

\section{2.~~THE FIRST STAR-FORMING SYSTEMS}

   Current cosmological models now provide us with a framework for
addressing the problem of early star formation and specifying plausible
initial conditions.  Recent progress in cosmology has led to a set
of variants of the standard CDM model which, while differing in
quantitative details, all make similar predictions about the way in
which structure emerged in the early universe.  In all of the currently
viable models, cosmic structure is built up hierarchically, and larger
systems are assembled from smaller ones by the accumulation of matter
at the nodes of a filamentary web-like network.  These models provide
a well-tested description of the development of galaxy clustering and
large-scale structure in the universe, but they do not yet predict
correctly the properties of individual galaxies and have not been tested
at all on much smaller scales; therefore we cannot yet be confident that
they predict in a quantitatively correct way the properties of the first
star-forming systems.  However, in all models, qualitatively similar
things are expected to happen on smaller scales at earlier times,
so we expect that the first star-forming systems were created in a
similar way by the accumulation of matter at the nodes of a filamentary
network.  The various currently viable models can then be used to make
extrapolations to smaller scales and earlier times that we can use as
working hypotheses to investigate early star formation.  We can also
hope that the physics of early star formation was simpler than that
of present-day star formation because the important physical processes
involved only various forms of hydrogen, and because turbulence and
magnetic fields might not yet have been introduced by the effects of
prior star formation.

   Current models predict that the first bound systems capable of
forming stars appeared during the first $10^8$ years of the history of
the universe at redshifts between 50 and 10, and that they had masses
between about $10^5$ and $10^8\,$M$_\odot$ (Peebles 1993; Haiman, Thoul,
\& Loeb 1996; Tegmark et al.\ 1997; Nishi \& Susa 1999; Miralda-Escud\'e
2000).  Most of this mass is dark matter, and the gas mass is about
an order of magnitude smaller. The predicted radii of these first
star-forming systems are between 10 and 500 parsecs, and their internal
velocity dispersions are between about 5 and 25 km/s.  These properties
are not greatly different from those of present-day star-forming regions
in galaxies, including giant molecular clouds, large complexes of gas
and young stars, and starburst regions.  However, an important
difference is that the temperatures of the first collapsing clouds must
have been much higher than those of present molecular clouds because of
the absence of any heavy elements to provide cooling.  In the metal-free
primordial clouds, the only possibility for cooling below $10^4\,$K is
provided by trace amounts of molecular hydrogen comprising up to about
$10^{-3}$ of the total hydrogen abundance, and H$_2$ molecules cannot
cool the gas significantly below 100$\,$K; the calculated temperatures
of primordial clouds are in fact mostly in the range 200--1000$\,$K
(Anninos \& Norman 1996; Haiman, Thoul, \& Loeb 1996; Tegmark et al.\
1997; Nakamura \& Umemura 1999; Abel, Bryan, \& Norman 1999; Bromm,
Coppi, \& Larson 1999).  Thus, thermal pressure must have played a
much more important role in primordial star formation than it does in
present-day star formation; in particular, the Jeans mass must have been
much larger in the primordial clouds than it is in present-day clouds,
since the densities of the first star-forming clouds were not very
different from those of present clouds, while their temperatures were
much higher.

\section{3.~~THERMAL PROPERTIES OF THE PRIMORDIAL GAS}

   Three different groups simulating the collapse and fragmentation
of primordial star-forming clouds have recently obtained consistent
results for their thermal behavior, and these results will be summarized
briefly here.  The most realistic calculations are those of Abel et al.\
(1998) and Abel, Bryan, \& Norman (1999; hereafter ABN), who have
started with a full cosmological simulation including a detailed
treatment of the gas physics, and have followed the evolution of the
first recollapsing density peak through many orders of magnitude in
density using a progressively finer grid.  At the latest stage reached,
this calculation is well on the way to modeling the formation of a
single massive star or small group of stars.  Somewhat more idealized
is the simulation by Bromm, Coppi, \& Larson (1999, 2000; hereafter BCL)
of the collapse of a simple `top-hat' cosmological density perturbation
with a standard spectrum of density fluctuations and a typical angular
momentum; this calculation uses an SPH technique intended to follow the
formation of a small group of dense clumps.  In this simulation the gas
collapses to a disk which then fragments into filaments and clumps, and
the collapse of the clumps is again followed through a large increase in
density.  The most idealized and least cosmology-dependent calculations
are those of Nakamura \& Umemura (1999, 2000; hereafter NU), who have
simulated the collapse and fragmentation of gas filaments with initial
densities and temperatures appropriate for primordial clouds; these
authors have followed the collapse of these filaments to higher
densities than the other groups and into the regime where opacity
becomes important.

   In the simulations of ABN and BCL, the initial recollapse of a
$3\sigma$ density peak at a redshift of $\sim\!30$ compresses the gas
and heats it to temperatures above 1000$\,$K; this in turn increases the
H$_2$ formation rate and causes the H$_2$ abundance to rise from its
initial value of about $10^{-6}$ to a quasi-equilibrium value of about
$10^{-3}$ of the total hydrogen abundance.  The additional molecular
hydrogen thus created cools the gas in the flattened disk-like
configuration resulting from the collapse to a temperature of about
200--300$\,$K, but the temperature then begins to rise again in the
dense clumps that form, reaching $\sim\!500$--$800\,$K at the highest
densities attained.  The simulations by NU of the fragmentation of
filaments do not start with any overall collapse and therefore do
not show the same initial rise and fall in temperature, but they
nevertheless yield similar H$_2$ abundances and similar temperatures
of $\sim\!300$--$500\,$K over a similar range of densities.  All three
sets of calculations converge into the same density-temperature regime
after the initial collapse has stopped and any flattened or filamentary
configuration produced by it is beginning to fragment into clumps; at
this stage, the density is about $10^3$--$10^4$ atoms per cm$^3$ in
all cases, and the temperature is about 200--300$\,$K.  Under these
conditions the Jeans mass is of the order of $10^3\,$M$_\odot$, and all
of these simulations find that clumps are formed with masses of this
order.

   Thus, while more work is needed to verify the generality of these
conclusions, it appears that the predicted thermal behavior and
fragmentation scale of primordial clouds are fairly robust results, and
depend mainly on the gas physics and not on the simulation techniques or
the cosmological model assumed.  An important feature of the gas physics
that may help to account for this convergence of results is that at
densities greater than $10^4\,$cm$^{-3}$ the level populations of the
H$_2$ molecule come into thermodynamic equilibrium, causing the density
dependence of the cooling rate to saturate and the cooling time to
become independent of density; as a result, the cooling time again
becomes longer than the free-fall time at the higher densities, and
the collapse is significantly slowed down from a free fall.  A further
effect that may also be relevant is that the gas is still confined
partly by the gravity of the dark matter when it begins to fragment
into clumps, so that it is not yet fully self-gravitating and lingers
for a time in this favored density-temperature regime, allowing more
time for the Jeans scale to be imprinted on the dynamics.

\section{4.~~FRAGMENTATION AND THE STELLAR MASS SCALE}

   During the collapse of the `top-hat' density perturbations studied by
BCL, the density fluctuations in the dark matter grow unimpeded, while
the gas at first retains a smoother distribution because it is warmer;
however, as the density of both the dark matter and the gas increase,
the gas increasingly responds to the dark matter density fluctuations
and begins to develop a similar clumpy structure.  After about a
free-fall time, the dark matter `virializes' to form a small dark halo,
while the gas settles into a smaller rotationally supported disk within
this halo.  For a system with a total mass of $2\times10^6\,$M$_\odot$
and a gas mass of $10^5\,$M$_\odot$ collapsing at a redshift of
$\sim\!30$, the radius of this disk is about 15~pc.  Irregularities in
the disk develop into filamentary spiral structure, and the filaments
then fragment into massive clumps.  The density fluctuations in the dark
matter thus appear to trigger the growth of structure in the gas and may
determine where the most massive clumps will form, but once the gas has
settled into a disk, its subsequent evolution appears to depend more on
the thermal properties of the disk than on the initial conditions for
the collapse.  In particular, the masses of the clumps depend mainly on
the Jeans scale in the disk and not on the nature of the initial density
fluctuations.  The most important effect of the dark matter may then
simply be that its gravity confines the gas in a disk long enough for
gravitational instabilities in the disk to play a significant role in
its evolution.

   The full cosmological simulation of ABN shows qualitatively similar
behavior, but it exhibits filamentary structure from the beginning
and yields a single dominant central clump in the first recollapsing
density peak.  The rezoning technique used by these authors follows with
ever-increasing resolution the collapse of this clump, and at the last
stage reached, it has developed a slowly contracting core with a mass of
about $100\,$M$_\odot$ surrounded by a flattened envelope with a mass of
about $1000\,$M$_\odot$ in which rotational support is important.  The
central core continues to contract nearly spherically and shows no sign
of fragmentation into smaller objects.  It is possible that additional
objects might form in the flattened disk-like region around it, but the
calculation focuses on resolving the collapse of the central core.  The
introduction of `sink particles' in the simulation of BCL allows the
calculation to be carried farther and the formation of a small group
of objects to be followed, at the expense of not resolving what happens
at the highest densities.  Again, the clumps formed show no sign of
fragmenting into smaller objects.  The masses of these clumps are
typically of the order of $10^3\,$M$_\odot$, with a range extending
from the resolution limit of $\sim\!10^2\,$M$_\odot$ to more than $10^4\,$M$_\odot$.  Experiments with a variety of initial conditions
suggest that a typical clump mass of $\approx\!10^3\,$M$_\odot$ is a
rather general result even when the overall structure of the system
becomes much more complex than the simple disk discussed above.

   The simulations of NU also find that primordial gas filaments tend
to fragment into clumps with masses of the order of $10^3\,$M$_\odot$.
If some of the gas condenses into much thinner and denser filaments
before fragmenting, objects with much smaller masses can also be formed.
The simulations do not yet indicate how much mass might fragment into
smaller objects in this way, but it seems unlikely that a major fraction
of the total mass will be involved.  A question of great interest is
whether any stars smaller than a solar mass can be formed; Uehara et
al.\ (1996) suggested that such stars cannot form under primordial
conditions because the onset of high opacity to the H$_2$ cooling
radiation sets a minimum fragment mass that is approximately the
Chandrasekhar mass, somewhat above $1\,$M$_\odot$.  NU confirm this
result from a more detailed treatment of the radiative transfer problem,
and they conclude that fragmentation can continue down to a minimum mass
between 1 and $2\,$M$_\odot$.  If there is indeed a minimum mass for
metal-free stars that is larger than $1\,$M$_\odot$, this would be a
very important result because it would imply that we should see no
metal-free stars at the present time, even if large numbers of such
stars had once been formed, since all of these stars should by now
have evolved.

\section{5.~~STAR FORMATION THEN AND NOW}

   It may be useful to compare early star formation with present-day
star formation in discussing the expected fragmentation scale and the
likely masses of the stars formed.  Since the predicted sizes and masses
of the first star-forming systems are not greatly different from those
of present-day molecular clouds, their average densities and internal
pressures are also not very different.  In fact, the typical gas
pressure in the disks discussed above, which have an average density of
$\sim\!10^3\,$cm$^{-3}$ and a temperature of $\sim\!300\,$K, is about
the same as the typical pressure in present-day cold molecular cloud
cores, which have a density of $\sim\!3\times10^4\,$cm$^{-3}$ and a
temperature of $10\,$K.  Thus, if one calculates the Jeans mass from
the temperature and pressure of a star-forming cloud, for example by
taking the mass of a marginally stable Bonnor-Ebert sphere which varies
as the square of the temperature and inversely as the square root of
the pressure, this mass is larger in primordial clouds than in present
clouds just by the square of the temperature.  Since the Jeans mass
in present-day molecular clouds is of the order of one solar mass
(see below), and since the temperature is about 30 times higher in
the primordial clouds, the Jeans mass in these clouds is predicted
to be about 1000 times higher, or about $10^3\,$M$_\odot$, as was noted
above. Note that the Jeans mass will remain higher than present values
even after the first heavy elements have been introduced, since the
temperature still cannot fall below the cosmic background temperature,
which is 57--85$\,$K at redshifts of 20--30; this is nearly an order
of magnitude higher than the present-day temperatures of cold molecular
cloud cores, implying a Jeans mass that is still almost two orders of
magnitude higher than present values, again assuming a similar pressure
(Larson 1998).

   How relevant is the Jeans mass in determining typical stellar masses?
This question has been controversial in recent years, and it has been
debated whether the present characteristic stellar mass of the order of
one solar mass is determined by the scale of cloud fragmentation or by
the onset of strong outflows that terminate the accretional growth of
protostars at some stage (Meyer et al.\ 2000).  Some recent evidence
suggests that stellar masses are closely related to the masses of the
dense clumps observed in star-forming molecular clouds, supporting the
view that the characteristic stellar mass is determined by the scale
of cloud fragmentation.  Motte, Andr\'e, \& Neri (1998) have observed
many small dense clumps in the $\rho$~Ophiuchus cloud that have masses
between 0.05 and $3\,$M$_\odot$, and they find that the mass spectrum of
these clumps is similar to the IMF of local field stars, including the
flattening below a solar mass that is indicative of a characteristic
mass around one solar mass.  The mass spectrum of these clumps has also
been compared with that of the pre-main-sequence stars in the same
cloud by Luhman \& Rieke (1999), and they find that the two functions
are indistinguishable from each other and from the IMF of the local
field stars.  This suggests that stars form with masses similar to
those of the observed dense clumps in molecular clouds, and that the
characteristic stellar mass is therefore determined by the typical
clump mass.  This mass is in turn found to be similar to the Jeans
mass calculated from the typical temperature and pressure in molecular
clouds, which is approximately one solar mass or slightly less (Larson
1985, 1996, 1999, 2000; Meyer et al.\ 2000).

   Since the predicted thermal behavior of primordial clouds is
qualitatively similar to that of present-day molecular clouds except
for temperatures that are about 30 times higher at a given pressure,
similar conclusions might be expected to hold for primordial clouds,
with masses about three orders of magnitude larger for both the clumps
and the stars formed.  This suggests that the first stars were very
massive objects, with typical masses of perhaps several hundred solar
masses.  However, definite conclusions cannot yet be drawn about the
masses of the first stars because the final fate of the clumps discussed
above has not yet been determined.  The possibility that they will
fragment into many smaller objects has not been ruled out, even though
this seems unlikely.  In the present-day case, it appears on both
observational and theoretical grounds that the fragmentation of
collapsing Jeans-mass clumps is limited to the formation of at most a
small multiple system, the typical outcome being the formation of a
binary system (Larson 1995).  Numerical simulations illustrating the
formation of binary and multiple systems (e.g., Burkert, Bate, \&
Bodenheimer 1997) have usually assumed an isothermal equation of state,
but the temperatures in the primordial clumps discussed above do not
remain constant but rise slowly as the clumps contract, and this
can only reduce the amount of subsequent fragmentation that occurs.
However, even if further fragmentation is unimportant, it remains
possible that significant differences from the typical present-day
situation could arise because of the much higher stellar masses expected
at early times; for example, radiative feedback effects such as the
dissociation of H$_2$ molecules by ultraviolet radiation might reduce
the efficiency of early star formation (Haiman 2000; Ferrara \& Ciardi
2000), and they might also reduce the masses of the stars formed by
preventing the accretion by them of most of the initial clump mass
(Abel 1999).

   In summary, it appears to be a fairly general result that the
first star-forming clouds fragment into massive clumps with masses of
the order of $10^3\,$M$_\odot$ and temperatures of a few hundred K;
this result depends mainly on the well-understood thermal physics of
the gas and not on the details of the initial conditions or the
simulation technique used.  It also appears unlikely that these massive
clumps will fragment into many smaller objects as they collapse to
higher densities.  The stars that form in them will then almost
certainly be much more massive than typical present-day stars.  The
IMF of the first stars was therefore almost certainly top-heavy,
and it seems very  unlikely that a standard present-day IMF could
have been produced.  The first stars might typically have been
massive ($\approx\!10^2\,$M$_\odot$) or possibly very massive
($\approx\!10^3\,$M$_\odot$) objects, perhaps similar in the latter
case to the `VMOs' studied by Carr, Bond, \& Arnett (1984) and Bond,
Arnett, \& Carr (1984) and reviewed by Carr (1994).  Of course, any
predictions concerning the properties of the first stars are as yet
untested by any direct observations, and we must await observations of
very high-redshift objects with instruments such as NGST before we can
know for sure whether the work that has been described here is on the
right track.

\section{6.~~POSSIBLE EFFECTS OF EARLY STAR FORMATION}

   How were the first star-forming systems related to the galaxies that
we presently see, and what role might the first stars have played in
accounting for the properties of the systems that we see?  The first
star-forming units probably cannot be identified with any presently
observed systems, since they would have been too small and too loosely
bound to survive or to retain any gas after forming the predicted
massive stars.  These stars would have evolved within a few Myr, and
any whose masses were larger than about 250$\,$M$_\odot$ would have
collapsed to black holes containing at least half of the initial stellar
mass (Bond, Arnett, \& Carr 1984; Heger, Woosley, \& Waters 2000).  If
a significant fraction of the first stars had such large masses, much
of the matter that condensed into them might soon have ended up in black
holes of similar mass.  Massive black holes formed in this way at early
times could have had a number of interesting consequences, including
seeding the formation of supermassive black holes in galactic nuclei
and thus accounting for the origin of AGNs.
 
   In standard hierarchical cosmologies, the first objects form
preferentially in the densest parts of the universe, and they then
become incorporated through a series of mergers into systems of larger
and larger size which eventually become present-day large galaxies and
clusters of galaxies.  The first $3\sigma$ density peaks are predicted
to be strongly clustered on the scale of clusters of galaxies and
significantly clustered even on the scale of individual galaxies,
and this means that the first stars or their remnants should now be concentrated in the inner parts of large galaxies, which in turn are
mostly in large clusters (Miralda-Escud\'e 2000; White \& Springel
2000).  That is, these objects should now be found mostly in places like
the inner parts of M87 rather than in the outer halos of galaxies like
the Milky Way.  If massive black holes were present from early times in
the dense regions that later became the inner parts of large galaxies,
they might have become increasingly centrally concentrated because of
the strong gravitational drag effects that would have been present in
such regions.  The most massive ones might then have served as the seeds
for building up larger black holes by accretion, and mergers among them
might also have contributed to building up very massive central black
holes.  If the present galaxies with large spheroids were built up by
a series of mergers of smaller systems that already contained central
black holes, these central black holes might have merged along with
their host systems to form increasingly massive black holes at the
centers of galaxies of increasing mass.  It is conceivable that most of
the remnants of the first stars could have ended up in this way in the
supermassive black holes of AGNs; in this case, an understanding of
early star formation could turn out to be very relevant to understanding
the origin of AGNs.

   The accumulation processes that build massive black holes in galactic
nuclei might be analogous to the processes that form massive stars at
the centers of star clusters.  Massive newly formed stars are always
found in clusters, typically near their centers, and this can only be
understood if these massive stars were in fact formed near the cluster
center (Bonnell \& Davies 1998).  Since the various accretion or
accumulation processes that might increase stellar masses are most
important in the dense central parts of forming clusters, while the
Jeans mass is not unusually high there, this suggests that massive stars
are built up by accumulation processes in clusters and are not formed
by direct cloud fragmentation (Larson 1982; Bonnell, Bate, \& Zinnecker
1998; Clarke, Bonnell, \& Hillenbrand 2000; Bonnell 2000).  The most
massive stars may even be produced by collisions and mergers between
less massive stars in the extremely dense cores of forming clusters
(Bonnell, Bate, \& Zinnecker 1998; Stahler, Palla, \& Ho 2000).  As is
the case with galaxies, interactions and mergers among subunits may play
an important role in driving accretion onto central objects and possibly
causing them to merge, and this could lead to the formation of stars of
increasing mass as clusters are built up by the merging of substructure
(Clarke, Bonnell, \& Hillenbrand 2000; Bonnell 2000; Larson 2000).  Such
processes might also account for the observed power-law upper stellar
IMF; in the simple model suggested by Larson (2000), the most massive
star accretes a fixed fraction (1/6) of the remaining gas each time two
subunits merge, and this leads to a Salpeter-like upper IMF with a slope
$x = 1.36$ in which the mass of the most massive star increases as the
0.74 power of the mass of the cluster.  A modification of this model
might be able to account for the fact that the masses of the nuclear
black holes in galaxies are approximately proportional to the bulge
mass, typically being about 0.005 times the bulge mass.  If nuclear
black holes are built up by the same kind of sequence of mergers and
associated accretion events, and if the resulting black holes merge into
a single object when their masses exceed $10^6\,$M$_\odot$, then the
central black hole mass increases in proportion to the bulge mass for
larger masses and is about 0.005 times the bulge mass, as observed.
Thus there could be a close analogy between black hole formation in
galactic nuclei and the formation of massive stars in clusters.

   The first stars must also have produced the first heavy elements
in the universe.  While stars more massive than 250$\,$M$_\odot$ are
predicted to collapse completely to black holes without ejecting any
heavy elements, somewhat less massive primordial stars would have
exploded as supernovae and begun to enrich their surroundings with
heavy elements.  Stars with masses in the range between about 100 and
$250\,$M$_\odot$ are predicted to be partly or completely disrupted by
the pair-production instability (Bond, Arnett, \& Carr 1984; Heger,
Woosley, \& Waters 2000), producing an energetic supernova event and
dispersing some or all of the heavy elements produced during their
evolution.  Thus such objects could plausibly have been the first
sources of heavy elements.  Metal-free stars with masses between about
35 and $100\,$M$_\odot$ probably again collapse to black holes (Heger,
Woosley, \& Waters 2000), while stars with masses between 10 and
$35\,$M$_\odot$ can explode as type II supernovae, possibly providing
a second source of heavy elements if significant numbers of such stars
were formed at the earliest times.

  The first star-forming systems were probably too short-lived and too
weakly bound to retain any of the heavy elements produced, so the first
systems capable of self-enrichment were probably larger systems that
formed somewhat later as larger cosmological structures collapsed,
incorporating some of the remnants and nucleosynthetic products of the
first systems.  The dwarf galaxies of the Local Group are found to have
a minimum mass of about $2\times10^7\,$M$_\odot$ (Mateo 2000), which
is about the minimum mass needed for a galaxy to retain ionized gas.
The retention of ionized gas is essential for subsequent star formation
and chemical enrichment to occur because most of the gas in a galaxy is
cycled many times through an ionized phase, and the dispersal and mixing
of heavy elements also probably occurs mostly in an ionized medium.
Preliminary calculations similar to those of BCL but with H$_2$ replaced
as the dominant coolant by a low abundance of heavy elements suggest
that in such circumstances there may be a threshold metallicity between
$10^{-4}$ and $10^{-3}$ times the solar value below which no cooling
occurs but above which some of the gas can cool to temperatures as
low as the cosmic background temperature.  This would reduce the
typical masses of the dense clumps formed, but not to present-day
values because the background temperature is still relatively high at
high redshifts.  For example, the first systems with masses as large
as $2\times10^7\,$M$_\odot$ are predicted to form at a redshift of $\sim\!25$, and the cosmic background temperature at that redshift is
$71\,$K.  If it is still valid to assume a pressure similar to that in
present-day star-forming clouds, the predicted Jeans mass is then about
$50\,$M$_\odot$, still much larger than present-day values, suggesting
that star formation at high redshifts would still have produced a
top-heavy IMF even after the first heavy elements had been introduced
(Larson 1998).

   Early star formation with a top-heavy IMF in those regions that later
became the inner parts of large galaxies could help to resolve a number
of problems regarding the chemical abundances of galaxies and quasars.
The heavy-element abundances in galaxies increase systematically with
mass up to the largest masses known, and they also increase radially
inward in large galaxies of all types.  Neither of these trends is fully
explained by standard models assuming a universal IMF, even if gas flows
are invoked, but both might be explained if early star formation with
a top-heavy IMF enriched preferentially the inner parts of the largest
galaxies.  The nuclear regions of the largest galaxies contain the most
metal-rich stars known, and quasars can be even more metal-rich, with
metallicities up to 5 or 10 times solar. These high metallicities cannot
be explained with a standard IMF, but they might be explainable if the
nuclear regions of galaxies containing AGNs were enriched by massive
stars formed at early times with a top-heavy IMF, as would be expected
in the picture sketched above.  Star formation with a top-heavy IMF
during the formation of cluster elliptical galaxies, perhaps associated
with merger-induced starbursts, could also help to explain the high
metallicities of both the stars and the hot gas in clusters of galaxies
(Zepf \& Silk 1996; Larson 1998).  As was noted by these authors, a
top-heavy early IMF might in addition help to account for the observed
far-infrared cosmic background radiation, which is believed to be
produced mostly by obscured high-redshift massive star formation,
without violating limits on the current density of low-mass stars in
the universe (see also Dwek et al.\ 1998).  Finally, we note that the
`G-dwarf problem' mentioned in the introduction might be solved or
alleviated if such processes made a significant contribution to
enriching galaxies generally, even outside the nuclear regions.

\section{7.~~SUMMARY}

   The first theoretical studies of the formation of primordial or
`population~III' stars based on current cosmological models with a full
treatment of the relevant physics have been begun within the past few
years, and work by several groups is continuing.  These studies have
obtained consistent results for the thermal behavior of the first
star-forming clouds, implying a scale of fragmentation that is of the
order of $10^3\,$M$_\odot$, and all of the simulations have found the
formation of clumps whose masses are of this order.  This strongly
suggests that the first stars were typically massive or very massive
objects, that is, it suggests a very top-heavy early stellar IMF,
although quantitative predictions are not yet possible.  More work
is needed to verify the generality of these results, and more work is
also needed to follow the later evolution of the clumps and predict the
masses of the stars that form in them, but it seems very unlikely that
the earliest star formation could have yielded a normal present-day IMF.
After the first massive stars had formed, the properties of star-forming
systems and the universe must rapidly have become much more complex as
many feedback effects came into play, but it appears that some of the
properties of galaxies and quasars might be explainable as a result of
early star formation with a top-heavy IMF occurring in the dense regions
that became the inner parts of large galaxies.  The coming years are
sure to see much progress, both theoretical and observational, in the
currently very active quest to understand early star formation.

\section{REFERENCES}

{\leftskip=5mm \parindent=-5mm

Abel T. 1999, private communication

Abel T., Anninos P., Norman M. L., Zhang, Y. 1998, ApJ 508, 518

Abel T., Bryan G., Norman M. L. 1999, in Evolution of Large-Scale
  Structure: From Recombination to Garching, Banday A.~J., Sheth R.~K.,
  Da Costa L.~N. (eds.), ESO, Garching, p.~363 (ABN)

Anninos P., Norman M. L. 1996, ApJ 460, 556

Bond J. R., Arnett W. D., Carr B. J. 1984, ApJ 280, 825

Bonnell I. A. 2000, in Stellar Clusters and Associations: Convection,
  Rotation, and Dynamos, in press (astro-ph/9908268)

Bonnell I. A., Davies M. B. 1998, MNRAS 295, 691

Bonnell I. A., Bate M. R., Zinnecker H. 1998, MNRAS 298, 93

Bromm V., Coppi P. A., Larson R. B. 1999, ApJ 527, L5 (BCL)

Bromm V., Coppi P. A., Larson R. B. 2000, in The First Stars, Weiss A.,
  Abel T., Hill V. (eds.), Springer, Berlin, in press

Burkert A., Bate M. R., Bodenheimer P. 1997, MNRAS 289, 497

Carr B. J. 1994, ARA\&A 32, 531

Carr B. J., Bond J. R., Arnett W. D. 1984, ApJ 277, 445

Clarke C. J., Bonnell I. A., Hillenbrand L. A. 2000, in Protostars
  and Planets IV, Mannings V., Boss A.~P., Russell S.~S. (eds.),
  University of Arizona Press, Tucson, in press (astro-ph/9903323)

Dwek E., Arendt R. G., Hauser M. G., Fixsen D., Kelsall T., Leisawitz
  D., Pei Y.~C., Wright E.~L., Mather J.~C., Moseley S.~H., Odegard N.,
  Shafer R., Silverberg R.~F., Weiland J.~L. 1998, ApJ, 508, 106

Ferrara A., Ciardi B. 2000, in The First Stars, Weiss A., Abel T., Hill
  V. (eds.), Springer, Berlin, in press

Haiman Z. 2000, in The First Stars, Weiss A., Abel T., Hill V. (eds.),
  Springer, Berlin, in press

Haiman Z., Loeb A. 1997, ApJ 483, 21

Haiman Z., Thoul A. A., Loeb A. 1996, ApJ 464, 523

Heger A., Woosley S. E., Waters R. 2000, in The First Stars, Weiss A.,
  Abel T., Hill V. (eds.), Springer, Berlin, in press

Larson R. B. 1982, MNRAS 200, 159

Larson R. B. 1985, MNRAS 214, 379

Larson R. B. 1995, MNRAS 272, 213

Larson R. B. 1996, in The Interplay Between Massive Star Formation, the
  ISM and Galaxy Evolution, Kunth D., Guiderdoni B., Heydari-Malayeri
  M., Thuan T. X. (eds.), Editions Fronti\`eres, Gif sur Yvette, p.~3

Larson R. B. 1998, MNRAS 301, 569

Larson R. B. 1999, in The Orion Complex Revisited, Mc\-Caughrean M.~J.,
  Burkert A. (eds.), ASP Conference Series, San Francisco, in press

Larson R. B. 2000, in Star Formation 1999, Nakamoto T. (ed.), in press
  (astro-ph/9908189)

Luhman K. L., Rieke G. H. 1999, ApJ 525, 440

Mateo M. 2000, in The First Stars, Weiss A., Abel T., Hill V. (eds.),
  Springer, Berlin, in press

Meyer M. R., Adams F. C., Hillenbrand L. A., Carpenter J. M., Larson
  R.~B. 2000, in Protostars and Planets IV, Mannings V., Boss A.~P.,
  Russell S.~S. (eds.), University of Arizona Press, Tucson, in press
  (astro-ph/9902198)

Miralda-Escud\'e J. 2000, in The First Stars, Weiss A., Abel T., Hill V.
  (eds.), Springer, Berlin, in press (astro-ph/9911214)

Motte F., Andr\'e P., Neri R. 1998, A\&A 336, 150

Nakamura F., Umemura M. 1999, ApJ 515, 239

Nakamura F., Umemura M. 2000, in The First Stars, Weiss A., Abel T.,
  Hill V. (eds.), Springer, Berlin, in press (NU)

Nishi R., Susa H. 1999, ApJ 523, L103

Peebles P. J. E. 1993, Principles of Physical Cosmology,  Princeton
  University Press, Princeton, p.~635

Stahler S. W., Palla F., \& Ho P. T. P. 2000, in Protostars and Planets
  IV, Mannings V., Boss A.~P., Russell S.~S. (eds.), University of
  Arizona Press, Tucson, in press

Tegmark M., Silk J., Rees M. J., Blanchard A., Abel T., Palla F. 1997,
  ApJ 474, 1

Uehara H., Susa H., Nishi R., Yamada M., Nakamura T. 1996, ApJ 473,
  L95

White S. D. M., Springel V. 2000, in The First Stars, Weiss A., Abel T.,
  Hill V. (eds.), Springer, Berlin, in press (astro-ph/9911378)

Zepf S. E., Silk J. 1996, ApJ, 466, 114

}
\bye